\documentclass[sigconf]{acmart}
\setcopyright{none}
\renewcommand\footnotetextcopyrightpermission[1]{} % removes footnote with conference information in first column
\usepackage{balance}
\usepackage{booktabs} % For formal tables
\usepackage{listings}
\usepackage{color}
\usepackage[skip=2pt]{caption}
\usepackage{subcaption}
\usepackage{cleveref}
\usepackage{textcomp}
\usepackage{graphicx}
\usepackage{algorithm,algpseudocode}

\usepackage{paralist}
\usepackage{amsthm}
\newtheorem{property}{Property}
    
\def\BibTeX{{\rm B\kern-.05em{\sc i\kern-.025em b}\kern-.08emT\kern-.1667em\lower.7ex\hbox{E}\kern-.125emX}}
\DeclareCaptionFormat{captionformat}{\fontsize{7}{8}\selectfont#1#2#3}
\definecolor{dkgreen}{rgb}{0,0.6,0}
\definecolor{gray}{rgb}{0.5,0.5,0.5}
\definecolor{mauve}{rgb}{0.58,0,0.82}

\mathchardef\mhyphen="2D

\begin{document}
\title{Learning To Rank Diversely At Airbnb}
\author{Malay Haldar,  Mustafa Abdool, Liwei He, Dillon Davis, Huiji Gao, Sanjeev Katariya}
\affiliation{%
  \institution{Airbnb, Inc.}
  \country{USA}
}
\email{malay.haldar@airbnb.com}

\renewcommand{\shortauthors}{Malay Haldar et al.}

\begin{abstract}
Airbnb is a two-sided marketplace, bringing together hosts who own listings for rent, with prospective guests from around the globe. Applying neural network--based learning to rank techniques has led to significant improvements in matching guests with hosts. These improvements in ranking were driven by a core strategy: order the listings by their estimated booking probabilities, then iterate on techniques to make these booking probability estimates more and more accurate. Embedded implicitly in this strategy was an assumption that the booking probability of a listing could be determined independently of other listings in search results. In this paper we discuss how this assumption, pervasive throughout the commonly-used learning to rank frameworks, is false. We provide a theoretical foundation correcting this assumption, followed by efficient neural network architectures based on the theory. Explicitly accounting for possible similarities between listings, and reducing them to diversify the search results generated strong positive impact. We discuss these metric wins as part of the online A/B tests of the theory. Our method provides a practical way to diversify search results for large-scale production ranking systems.
\end{abstract}




\maketitle
\pagestyle{plain}
\pagenumbering{gobble}

\section{Introduction}
Production e-commerce search ranking systems have to account for a range of target metrics, and search ranking at Airbnb is no exception. Ranking at Airbnb aims to optimize the guest and host experience end-to-end. Targets include increasing bookings, decreasing negative outcomes like cancellations, as well as increasing final trip ratings. Each of these targets are valuable and different parts of ranking are dedicated towards improving these targets. The core model that forms the foundation of ranking is focused on increasing bookings, and it does so by ordering listings by their booking probability. 

Formulating ranking as the ordering of listings by their booking probability was the result of a long evolutionary process. Prior to 2015, listings at Airbnb were ranked by scoring functions designed by engineers, which were related to booking probability only indirectly. The scoring functions took into account important attributes of the listings such as price, location, and review ratings. These attributes of the listings were studied to infer their relation to bookings, and scoring functions were designed to uprank listings with preferred attributes. This process was largely automated in 2015 with the launch of a gradient-boosted decision tree ($GBDT$) model. The $GBDT$ was a regression model, targeting a utility score assigned to each listing based on past user interactions. For example, a booking could be assigned a utility of $1.0$, a click on the listing a utility of $0.2$, an impression $0.0$, while a cancellation a negative utility of $-0.5$. Like the manually crafted scoring functions, the $GBDT$ model correlated with booking probability, but the association still remained indirect.

Next in the step of evolution came neural networks. Our journey is described in ~\cite{kdd19} and ~\cite{kdd20}. Casting the modeling task as pairwise learning to rank, optimizing it using cross-entropy loss, and using normalized discounted cumulative gain ($NDCG$) for evaluation brought booking probability into sharp focus. Over multiple iterations, we pushed the accuracy of the booking probability prediction, resulting in significant gains in bookings online. The baseline now stands at a formidable level. For a lot of applications, reaching performance comparable to humans is the gold standard. In our case, the model has far surpassed this level. As part of an evaluation exercise, a task was given to ranking engineers to identify which listing out of a pair was booked by a searcher. Ranking engineers could identify the booked listing for 70\% of the pairs, in comparison to 88\% for the ranking model.

To further improve the model, it was time to look beyond the success of the pairwise learning to rank approach. In our pursuit to improve bookings and $NDCG$, focus turned to the question: what aspects of the ranking model {\it could not} be improved by refining the pairwise probability of booking? One area that came up as a possible answer was, diversity in search ranking -- or the lack thereof.

\section{Why search results lack diversity?} \label{conehead}
To understand why typical ranking solutions do not diversify search results out of the box, we dive deeper into the specific case of pairwise learning to rank. A similar reasoning applies to pointwise learning to rank. The discussion around listwise learning to rank is more involved, which we visit in a later section.

\subsection*{How do listings get ranked?}
In pairwise learning to rank, we construct training examples from pairs of listings. Consider a search result in response to query $q$, issued by a user $u$. Let $l_x$ be a listing in the search result that was booked, and $l_y$ a listing that appeared along with $l_x$ but was not booked. Then the pair $\{l_x, l_y\}$ forms a training example. Let $f_{\theta}(q, u, l)$ represent a model with query, user, and the listing to be ranked as inputs, and $\theta$ the trainable parameters, or “weights” of the model.

To train the model, we get the estimated logits for both the listings in a training example as
\begin{equation*}
 \text{logit}_x=f_{\theta}(q, u, l_x)\; ; \; \text{logit}_y=f_{\theta}(q, u, l_y)
\end{equation*}
 The cross-entropy loss is given by
 \begin{equation}\label{eq:crossentropy}
  \text{crossEntropy} = -\log \left(\frac{e^{\text{logit}_x}}{e^{\text{logit}_x}+e^{\text{logit}_y}} \right)
\end{equation}
An example implementation of this loss in TensorFlow\texttrademark:
{\fontfamily{inconsolata}\selectfont
\begin{lstlisting}[basicstyle=\small]
import tensorflow as tf
def get_xentropy(logits_booked, logits_not_booked):
  logit_diffs = logits_booked - logits_not_booked
   xentropy = tf.nn.sigmoid_cross_entropy_with_logits(
      labels=tf.ones_like(logit_diffs),
      logits=logit_diffs)
   return tf.reduce_mean(xentropy)
\end{lstlisting}}
Minimizing the cross-entropy loss summed over all the training examples leads to a model where for any given pair of listings, $e^{\text{logit}_x}/(e^{\text{logit}_x}+e^{\text{logit}_y})$ can be interpreted as the pairwise booking probability, written as $P_{\text{booking}}(l_x>l_y \mid q, u)$. We omit the conditional $\{q, u\}$ going forward for brevity, and write it as $P_{\text{booking}}(l_x>l_y)$. If the combined booking count for $l_x$ and $l_y$ is scaled to $1.0$, then  $P_{\text{booking}}(l_x>l_y)$ represents the estimated fraction of bookings for $l_x$, whereas $(1 -P_{\text{booking}}(l_x>l_y))$ represents the fraction for $l_y$.

If we define $P_{\text{booking}}(l_x)$ as the ordinary pointwise probability of booking $l_x$ given an impression in search results, and similarly define $P_{\text{booking}}(l_y)$, then they can be related to the pairwise booking probability by the Bradley–Terry model~\cite{wiki:bradleyterry}:
\begin{equation*}
P_{\text{booking}}(l_x>l_y)=\frac{P_{\text{booking}}(l_x)}{P_{\text{booking}}(l_x)+P_{\text{booking}}(l_y)}
\end{equation*}
For example, consider $100$ search results where $l_x$ and $l_y$ were shown. If $l_x$ got $2$ bookings while $l_y$ got $6$, then
\begin{align*}
 P_{\text{booking}}(l_x) =\frac{2}{100}\: &;\: P_{\text{booking}}(l_y)=\frac{6}{100} \\
 P_{\text{booking}}(l_x>l_y)&=\frac{2}{2+6} =0.25
\end{align*} 
Given the pair $\{l_x, l_y\}$, for every $1$ booking of $l_x$, we expect $3$ of $l_y$.

In order to rank a given set of listings $\{l_a, l_b, ..., l_z\}$, we apply the ranking model $f_{\theta}(q, u, l)$ to each listing to get the corresponding logits $\{f_{\theta}(q, u, l_a), f_{\theta}(q, u, l_b),.., f_{\theta}(q, u, l_z)\}$, then sort the listings by their logits in descending order.

When sorting by logits, the condition for ranking listing $l_x$ higher than $l_y$ can be successively restated as:
\begin{align} \label{pinequality}
\begin{split}
f_{\theta}(q, u, l_x) &>f_{\theta}(q, u, l_y) \\
e^{f_{\theta}(q, u, l_x)} &>e^{f_{\theta}(q, u, l_y)} \\
\frac{e^{f_{\theta}(q, u, l_x)}}{e^{f_{\theta}(q, u, l_x)} + e^{f_{\theta}(q, u, l_y)}} &> \frac{e^{f_{\theta}(q, u, l_y)}}{e^{f_{\theta}(q, u, l_x)} + e^{f_{\theta}(q, u, l_y)}} \\
P_{\text{booking}}(l_x>l_y) &> P_{\text{booking}}(l_y>l_x)
\end{split}
\end{align}
We can therefore claim that pairwise learning to rank orders the listings by their pairwise booking probabilities. Inequality~\ref{pinequality} can be further rewritten as:
\begin{alignat*}{2}
P_{\text{booking}}(l_x>l_y) &>P_{\text{booking}}(l_y>l_x) \\
\frac{P_{\text{booking}}(l_x)}{P_{\text{booking}}(l_x)+P_{\text{booking}}(l_y)} &>\frac{P_{\text{booking}}(l_y)}{P_{\text{booking}}(l_x)+P_{\text{booking}}(l_y)} \\
P_{\text{booking}}(l_x) &>P_{\text{booking}}(l_y)
\end{alignat*}
This establishes the following:
\begin{property} \label{prop1}
Ranking listings by their pairwise booking logits is equivalent to ranking them by their pointwise booking probabilities.
\end{property}

\subsection*{How does ranking by booking probability affect diversity?}
Inequality~\ref{pinequality} can also be expressed as:
\begin{alignat*}{2}
P_{\text{booking}}(l_x>l_y) &>P_{\text{booking}}(l_y>l_x) \\
P_{\text{booking}}(l_x>l_y) &>1-P_{\text{booking}}(l_x>l_y) \\
P_{\text{booking}}(l_x>l_y) &>0.5
\end{alignat*}
which establishes:
\begin{property} \label{prop2}
When ranking two listings by their pairwise booking logits, the one estimated to get more than $50\%$ of the bookings is ranked higher.
\end{property}
At Airbnb, a query $q$ from user $u$ typically results in $0$ or $1$ booking. In such a scenario, the listing estimated to get more than $50\%$ of the bookings also represents the listing preferred by more than $50\%$ of the past bookers who issued query $q$, or the {\it majority preference} for the segment $\{q, u\}$. It is built into the very foundation of pairwise learning to rank to abide by the majority preference at each ranking position. Note, as in real life elections, majority is defined by those who vote, or in our case, those who book. The majority preference is not defined by the entire population that visits Airbnb, as preference of non-bookers remain hidden. In theory, a heavily-personalized ranking model could customize results for the minority preference appropriately. In practice though, accurately identifying users with minority preferences, and personalizing their search results in a fruitful manner, is an open challenge at Airbnb. The majority preference dominates the model's decisions for all practical purposes.

\begin{figure}
\includegraphics[height=1.25in, width=2.5in]{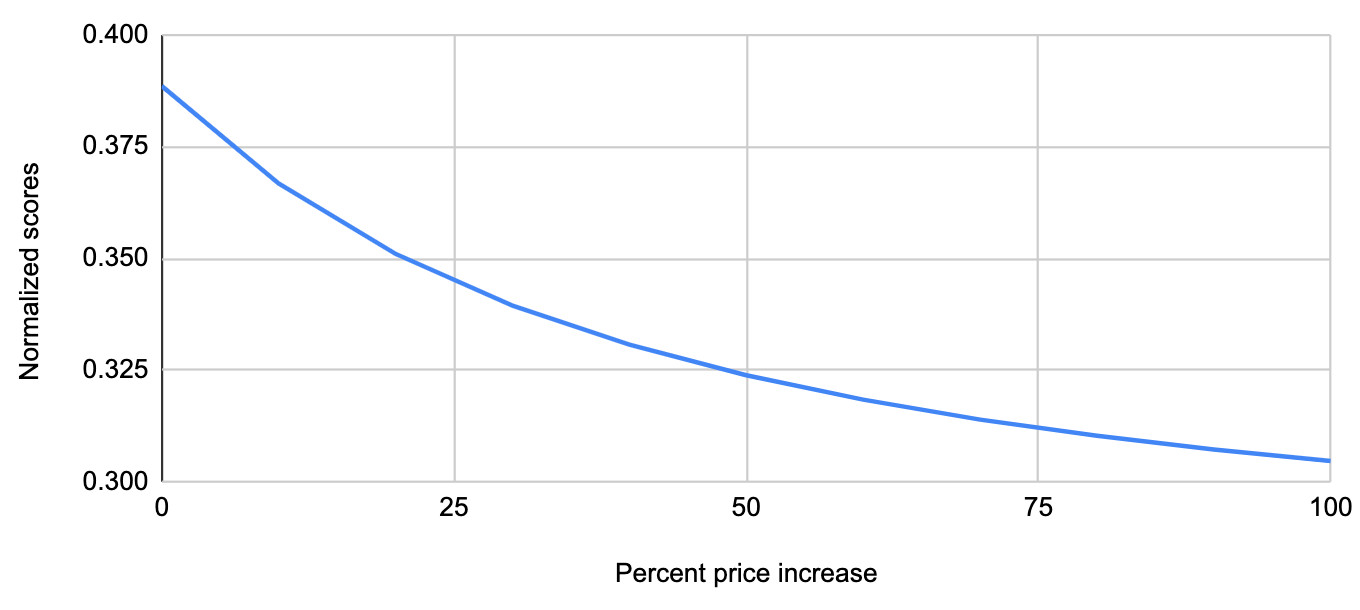}
\caption{\textmd{X-axis: percent increase in price. Y-axis: model scores normalized per query. Plot shows the average model score changes over a random sample of 100K listings.}}
\label{fig:pricecurve}
\end{figure}

But majority preference isn’t necessarily the best way to accommodate the preference of the entire population. Let’s elaborate the idea by an example. An important consideration for bookers at Airbnb is price, and the majority of bookings lean towards economical ones. Learning from this user behavior, the ranking model demotes listings if price increases. Figure~\ref{fig:pricecurve} shows how the normalized model score for a listing decreases as we increase price along the x-axis.

\begin{figure}
\includegraphics[height=1.8in, width=3in]{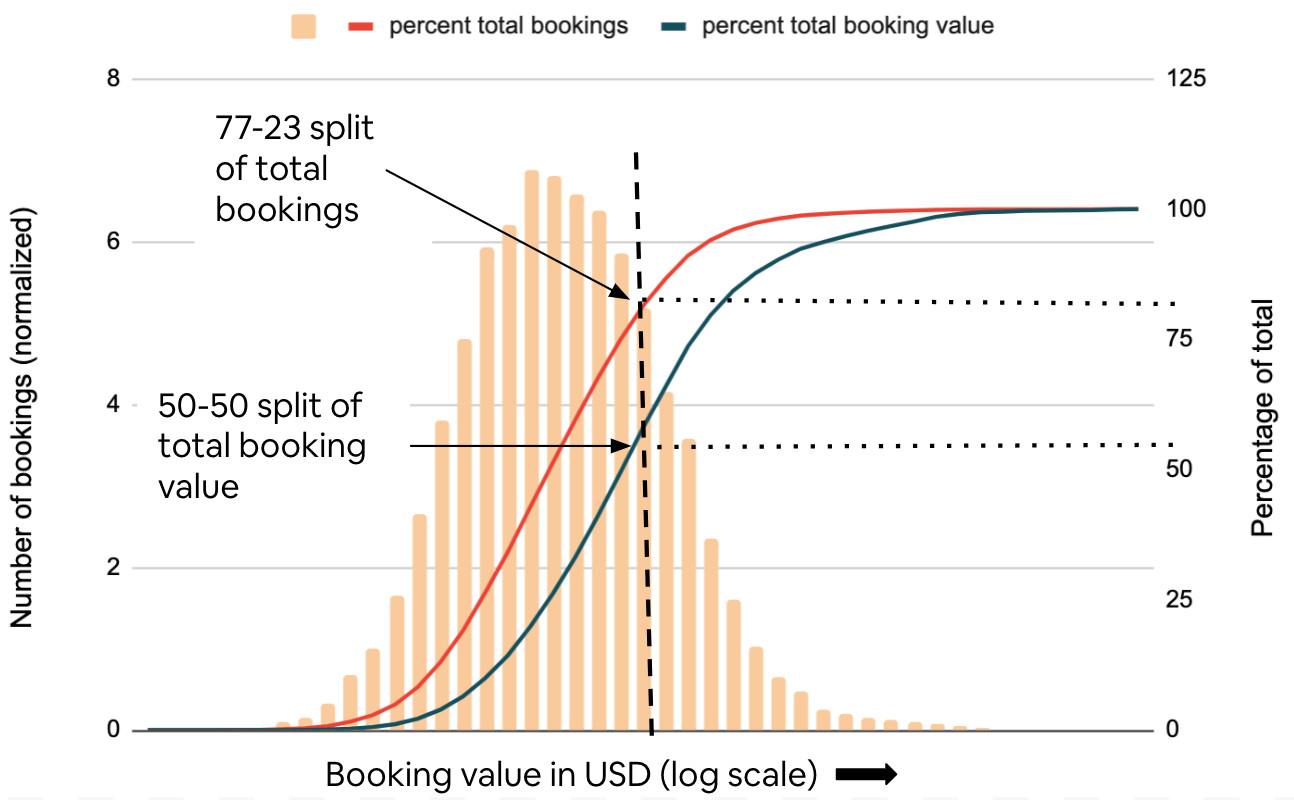}
\caption{\textmd{A distribution of booking values for 2 guests, 2 nights bookings in Rome. X-axis corresponds to booking values in USD, log-scale. Left y-axis is the number of bookings corresponding to each price point on the x-axis. The orange shape confirms the log-normal distribution of booking value. Red line plots the percentage of total bookings in Rome that have booking value less than or equal to the corresponding point on x-axis, and the green line plots the percentage of total booking value for Rome covered by those bookings. Splitting total booking value 50/50 splits bookings into two unequal groups of 80/20.}}
\label{fig:paretosplit}
\end{figure}

Experiments directly measuring price sensitivity confirm the same, that bookings drop sharply in response to price increases. But gravity towards affordability doesn’t tell the whole story, as Airbnb is a very diverse marketplace. The pareto principle ~\cite{wiki:pareto}, or the $80/20$ rule, provides a better perspective. In most cities, if we segment the total value of bookings, roughly ${\sim}20\%$ of bookings account for $50\%$ of the aggregated booking value (see Figure~\ref{fig:paretosplit}).

Using booking value as an indicator of quality, we can apply the broad classification that the  majority ${\sim}80\%$ of users are affordability-leaning. The remaining minority ${\sim}20\%$ are quality-leaning, with average booking value four times higher compared to the $80\%$ majority. In reality, though, users can't be boxed into such neat binary classifications. Every guest has preference towards both quality and affordability, and the distribution is continuous. Still, the simplified and binarized 80/20 perspective is useful to see where ranking is falling short.

Let’s consider how the first page of search results, where most of the user attention is spent, is impacted by the two effects:
\begin{compactitem}
\item {\it Majority principle}: ranking is driven by the majority preference at each position, as shown in Property~\ref{prop2}.
\item {\it Pareto principle}: user preferences are distributed smoothly with a long tail, and can be roughly binarized by a 80/20 split, as shown in Figure~\ref{fig:paretosplit}.
\end{compactitem}
The combined effect is that $100\%$ of the first page results get dictated by what the $80\%$ majority prefers. It’s the tyranny of the majority ~\cite{wiki:tyranny} in search ranking.

Intuitively, this suggests that factoring in the minority preference, and giving them proportionate representation, should improve the overall utility of search results. At the same time, the model based on pairwise learning to rank, friend of the tyrant majority oblivious of diversity, was the undisputed champion when it came to delivering bookings. That led us to ask a few questions:
\begin{compactitem}
\item The core focus of ranking is to optimize for total bookings. Is improving diversity of search results a distraction from the core focus?
\item A lot of effort has gone into refining the pairwise booking probability model. Can some simple changes allow it to tackle diversity "for free”?
\item Evaluating $NDCG$ offline, paired with measuring bookings in A/B tests online, is the gold standard for evaluating ranking models. Does diversity need a radically different evaluation method?
\end{compactitem}
The answers to the questions happen to be -- No, No, and No. We dive into further details in the next section.

\section{Can improving diversity drive bookings gain?}
To see how diversification of search results can lead to bookings gain, we start by examining how improving $NDCG$ leads to increased bookings. We then reason that optimal diversity in search results should improve $NDCG$, thereby improving bookings. Finally, we look at the relation between pairwise booking probabilities and $NDCG$, and the unique aspects of the $NDCG$ gain from diversity optimization.

\subsection*{Why does NDCG correlate with total bookings?}
A useful tool for our discussion in this section is the game of roulette. Consider $N$ games being played over the course of an evening. In each game, we have bets on $K$ different numbers. Let $P_{\text{win}}(i,j)$ refer to the probability of win on the $j$th bet in the $i$th game. Let $win(i, j)$ be the corresponding winning amount in dollars. The total dollars won is expected to be:
\begin{equation}
E[\text{total dollars won}]=\sum\limits_{i=0}^{N}\sum\limits_{j=0}^{K}P_{\text{win}}(i, j)*win(i, j)
\end{equation}
At the end of the evening, we can sum up the actual realized wins from each bet to get the total, which is the observed outcome of the stochastic process. The {\it observed total dollars won} would converge to the {\it expected total dollars won}, provided the number of games played $N$ is large enough~\cite{wiki:expectedvalue}. 

To analyze the total number of bookings from a given set of search results, we can employ a reasoning similar to roulette. Consider $N$ search results, each with $K$ listings. Getting a booking for a listing is equivalent to the winning event in roulette, and it depends on the attributes of the listing, as well as the listing’s position in search results. For the $i$th search result, let the listing placed at the $j$th position be $l_{i,j}$. We denote its complete probability of booking as $P_{\text{booking}}(l_{i,j})*P_{\text{attention}}(j)$, where $P_{\text{booking}}(l_{i,j})$ is the booking probability of the listing $l_{i,j}$, and $P_{\text{attention}}(j)$ is the probability that the guest examines the $j$th position of the search result. The equivalent of the winning amount in each case is one booking, so $win(i,j)=1$. The total number of bookings expected can therefore be written as:
\begin{equation}\label{eq:expected_bookings}
E[\text{bookings}]=\sum\limits_{i=0} ^{N}\sum\limits_{j=0}^{K}P_{\text{booking}}(l_{i,j})*P_{\text{attention}}(j)
\end{equation}

Under the assumption that user attention drops monotonically as they scan the search results from top to bottom, i.e, $P_{\text{attention}}(a)>P_{\text{attention}}(b)$ if $a < b$, we can show that $E[\text{bookings}]$ is maximized if the listings are sorted by their booking probabilities. This property can be established using a proof by contradiction.

Assume we have maximized $E[\text{bookings}]$, but the listings are {\it not} sorted by their booking probabilities. Then there must exist a pair of listings such that $P_{\text{booking}}(l_{i,x}) <P_{\text{booking}}(l_{i,y})$ and $P_{\text{attention}}(x)>P_{\text{attention}}(y)$. Consider swapping the positions of the two listings $l_{i,x}$ and $l_{i,y}$. The difference in expected bookings due to the swap is given by:
\begin{align}\label{eq:swapbookings}
\begin{split}
{\mathsf B}_{x} = P_{\text{booking}}(&l_{i,x}) \; ; {\mathsf B}_{y} = P_{\text{booking}}(l_{i, y}) \; ; \; {\mathsf B}_{x} < {\mathsf B}_{y}\\
{\mathsf A}_{x} = P_{\text{attention}}&(x) \; ; \; {\mathsf A}_{y} = P_{\text{attention}}(y) \; ; \;  {\mathsf A}_{x}  > {\mathsf A}_{y}\\
\Delta E[\text{bookings}] = & \;({\mathsf B}_{x} {\mathsf A}_{y} +{\mathsf B}_{y}{\mathsf A}_{x}) - ({\mathsf B}_{x}{\mathsf A}_{x}+{\mathsf B}_{y}{\mathsf A}_{y}) \\
   = & \;({\mathsf B}_{y} -{\mathsf B}_{x})({\mathsf A}_{x} - {\mathsf A}_{y} )
\end{split}
\end{align}

Since both terms of the product in Equation~\ref{eq:swapbookings} are positive, $\Delta E[\text{bookings}] > 0$. This implies the previous sum could not be the maximum. Hence the listings must be sorted by booking probability to maximize the total expected bookings. For alternate arguments supporting the property, see ~\cite{probabilityprinciple}.

The difference in total expected bookings due to an unsorted pair, $\Delta E[\text{bookings}]$, is the product of the difference in booking probability and the difference in attention to the two positions. As we discuss next, $NDCG$ tracks total bookings so well because it follows the same conditions.

Let’s begin by considering how $NDCG$ is computed. We adopt a binary definition of relevance where a booked listing has a relevance of $1$, and all other listings $0$ relevance. For simplicity, we assume only a single listing is booked from a given search result. $NDCG$ can then be written as:
\begin{equation}\label{eq:ndcg}
NDCG = \frac{1}{N}\sum\limits_{i=0}^{N}\frac{\text{log}(2)}{\text{log}(2 + \text{pos}_{i})}
\end{equation}
where $\text{pos}_{i}$ refers to the position of the booked listing in the $i$th search.

We can map the computation of $NDCG$ back to our game of roulette, where we consider each search result as an individual game as before. The winning event is defined as a listing getting booked, and the winning amount at position $j$ is set to $\text{log}(2)/(\text{log}(2 + j))$. The formula for $NDCG$ in Equation~\ref{eq:ndcg} is computing the observed total value of this stochastic process, a simple sum over the realized individual wins. What about the expected value of $NDCG$? The expected value would be the sum of the winning amounts for each position, weighted by the probability of attaining the win. This can be written as:
\begin{equation} \label{eq:2}
 E[NDCG]=\frac{1}{N}\sum\limits_{i=0}^{N}\sum\limits_{j=0}^{K}P(l_{i,j})*\frac{\text{log}(2)}{\text{log}(2 + j)} 
\end{equation}
Provided we evaluate $NDCG$ over a large number of searches, the expected and observed values would converge, and maximizing Equation~\ref{eq:ndcg} is equivalent to maximizing Equation~\ref{eq:2}.

The monotonic decay of $\text{log}(2)/\text{log}(2 + j)$ together with Equation~\ref{eq:2}  implies that sorting the listings by their booking probabilities maximizes $NDCG$. The proof is by contradiction once again. Further, the drop in $NDCG$ from a pair of listings $l_{i,x}$ and $l_{i,y}$ not ordered by booking probabilities is given by:
\begin{align}\label{eq:swapndcg}
\begin{split}
 \Delta E[NDCG] = & (P_{\text{booking}}(l_{i,y})-P_{\text{booking}}(l_{i,x}))\\
 &* \left(\frac{\text{log}(2)}{\text{log}(2+x)}- \frac{\text{log}(2)}{\text{log}(2+y)}\right)
 \end{split}
\end{align}

$\Delta E[NDCG]$ in Equation~\ref{eq:swapndcg} correlates with $\Delta E[\text{bookings}]$ in Equation~\ref{eq:swapbookings} because the positional discount curve is constructed based on how user attention decays by position, making $\text{log}(2)/\text{log}(2+x)$ proportional to $P_{\text{attention}}(x)$. As a result, a gain in $NDCG$ is also a strong indicator of gain in total bookings.

To recap, both total bookings and $NDCG$ are maximized when listings are sorted by their booking probabilities. In the case of total bookings, this arises due to the fact that we accurately represented the probability of booking as $P_{\text{booking}}(l_{i,j})*P_{\text{attention}}(j)$. However for $NDCG$, we made the simplified assumption that the probability of booking is given by $P_{\text{booking}}(l_{i,j})$ alone, and independent of the position where the listing is placed. This assumption is a useful one, since 
it allows us to shuffle listings around in offline analysis, and compute $NDCG$ without requiring fresh user input. But then, to align $NDCG$ with total bookings, we need to weigh each potential booking by its positional discount.

\subsection*{Can NDCG measure the impact of diversity?}
Utilizing the fact that listings ordered by their booking probabilities maximize $NDCG$ and total bookings, we can design a straightforward iterative algorithm to construct optimal search results as shown in Algorithm~\ref{algo1}.
\begin{algorithm}
\caption{Ranking by booking probabilities} \label{algo1}
\begin{algorithmic}[1]
\Require{A set of $N$ listings $\{l_0, l_1, \dots l_{N-1}\}$}
\Ensure{Listing positions $\{\text{pos}(l_0), \text{pos}(l_1), \dots \text{pos}(l_{N-1})\}$}
\Statex
\State $\mathcal{L}  \gets \{l_0, l_1 \dots l_{N-1}\}$
\For {$k \gets 0$ until $N$}
    \State Compute $\text{logit}(l_i) \gets f_{\theta}(q, u, l_i)$ foreach $l_i \in \mathcal{L}$
    \State $l_{max} \gets \text{argmax}(\text{logit}(l_i), l_i  \in \mathcal{L})$
    \State $\text{pos}(l_{max}) \gets k$
    \State $\mathcal{L} \gets \mathcal{L} \setminus l_{max}$
\EndFor
\end{algorithmic}
\end{algorithm}

In Algorithm~\ref{algo1}, we rely on Property~\ref{prop1}, which allows us to use the pairwise booking logit interchangeably with booking probability. $\text{logit}(l_i)$ on line 3 depends only on the attributes of the listing being ranked, $l_i$, besides the query and the user. Since attributes of a listing are invariant all throughout, line 3 in Algorithm~\ref{algo1} can be taken out of the for loop. This reduces Algorithm~\ref{algo1} to computing $\text{logit}(l_i)$ once for each listing, and then using it to sort the listings.

But it is {\it not a binding restriction} that the booking probability of a listing should be independent of the other listings, and must depend on attributes of the given listing alone. Specifically, consider iteration $K+1$ of the for loop in Algorithm~\ref{algo1}. We have already placed listings in position $0$ through $K-1$, and we need to select a listing for position $K$ by computing the logits of the remaining $N-K$ listings. When computing the logit for a given listing $l_i$, we can factor in the attributes of $l_i$, {\it as well as all the attributes of the $K$ listings placed at $0$ through $K-1$}. By including the attributes of listings at $0$ through $K-1$, we can ensure the calculated logits of the $N-K$ listings are more accurate. Listings being ranked which are too similar to the listings at $0$ through $K-1$ can have their logits corrected to lower values.

Maximization of $NDCG$ is preserved through this process of extending the inputs to include attributes of the listings placed at $0$ through $K-1$. The critical step in the proof for maximal $NDCG$, where we swap the listings not ordered by booking probability to demonstrate a contradiction, continues to work. That's because the extended inputs from the listings at $0$ through $K-1$ are invariant in the swapping process.

This provides a mechanism for diversification that is aligned with maximizing $NDCG$. We require no new metrics to evaluate diversity. Due to the relation between $NDCG$ and total bookings, we expect this mechanism to directly increase total bookings as well. To summarize, diversity removes redundant choices, thereby improving utilization of positions in search results, which get reflected in improved $NDCG$ and total bookings.

While we don’t need a new metric for {\it evaluating} diversity, the situation is different when it comes to {\it implementing} diversity. For diversity aware booking probabilities, we require the attributes of the listings that are placed before. This information is not available in the garden variety pointwise or pairwise learning to rank frameworks. Hence we need a new kind of model, which we discuss next.

\section{How to implement diversity in ranking?}\label{crt}
In this section we build a framework to rank listings for $N$ positions while incorporating diversity. Instead of building a single model, we build $N$ models, a dedicated model for each of the $N$ positions.

For position $0$, we reuse the regular pairwise booking probability model described in Section~\ref{conehead}. Let’s name the model $f_{0, \theta_0}(q, u, l)$, where the $0$ index refers to the position in search result, and $\theta_0$ the parameters of the model. To recap from Section~\ref{conehead}, $f_{0, \theta_0}(q, u, l)$ maps each listing to a pairwise logit, where sigmoid of the difference of two pairwise logits gives the pairwise booking probability.

Now let’s construct $f_{1, \theta_1}(q, u, l, l_0)$, the model for position $1$. This model has an additional input $l_0$, which we call the \textbf{\textit{antecedent listing}} from position $0$. A user scanning the search results from top to bottom would consider the listing at position $1$, only if the antecedent listing at position $0$ did not meet their requirements. See ~\cite{clickmodel1} and ~\cite{clickmodel2} for an in-depth study of this phenomenon. Thus when ranking for position $1$, we have incrementally more information than we did when ranking for position $0$. Leveraging this new information, we construct $f_{1, \theta_1}(q, u, l, l_0)$ conditional upon the fact that the user has rejected the listing at position $0$.

\begin{figure}
\includegraphics[height=1.9in, width=3.1in]{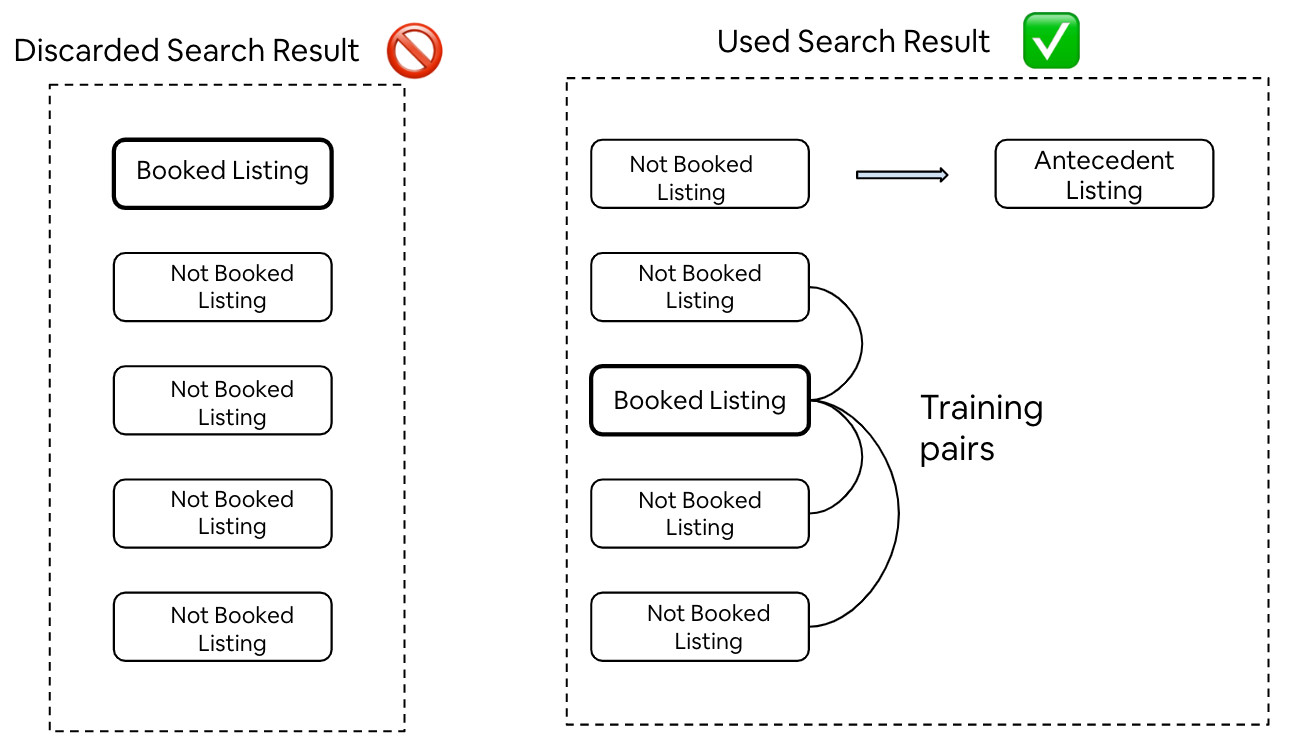}
\caption{\textmd{Training data construction for $f_{1, \theta_1}(q, u, l, l_0)$}}
\label{fig:simtrain}
\end{figure}

To construct training examples for this conditional pairwise booking probability model, we go through the search logs and {\it discard all searches where the listing at position $0$ was booked}. For the remaining searches, we set aside the listing at position $0$, denoting it the antecedent listing. From the listings below position $0$, we create pairs of booked and not booked listings, similar to how pairwise booking examples are created. Figure ~\ref{fig:simtrain} illustrates this. 

For a training example with $l_x$ as the booked listing, $l_y$ as the not booked listing, and $l_0$ as the antecedent listing,  we compute
\begin{equation*}
 \text{logit}_x=f_{1, \theta_1}(q, u, l_x, l_0) ; \text{logit}_y=f_{1, \theta_1}(q, u, l_y, l_0)
\end{equation*}
 and cross-entropy loss the same as Equation~\ref{eq:crossentropy}. By minimizing the cross-entropy loss summed over all training examples, we can infer parameters $\theta_1$ of the model such that $e^{\text{logit}_x}/(e^{\text{logit}_x}+e^{\text{logit}_y})$ represents the pairwise probability a user will book $l_x$ over $l_y$, {\it given the condition they have rejected the antecedent listing $l_0$}. We write this conditional probability as $P_{\text{booking}}(l_x>l_y \mid \mathcal{A}=\{l_0\})$. The conditional part of this pairwise booking probability arises because of the training data. The model is learning about pairwise preference between $l_x$ vs. $l_y$, but only from those users who have rejected $l_0$. Intuitively, we expect the choice of such users to be different from $l_0$, providing us a notion of diversity that can be learnt from the training data. 

For position $2$, we follow a similar strategy. We discard all searches where either listing at position $0$ or $1$ was booked, to arrive at the model $f_{2, \theta_2}(q, u, l, l_0, l_1)$ which gives $P_{\text{booking}}(l_x>l_y \mid \mathcal{A}=\{l_0, l_1\})$. Generalizing for position $k$, we get $f_{k, \theta_{k}}(q, u, l, l_{0 \rightarrow k-1})$ which predicts the conditional probability $P_{\text{booking}}(l_x>l_y \mid \mathcal{A}=\{l_{0 \rightarrow k-1}\})$. 

For ranking $N$ positions, we now have $N$ distinct ranking models $\{f_{0, \theta_0}(q, u, l), f_{1, \theta_1}(q, u, l, l_0), \dots, f_{N-1, \theta_{N-1}}(q, u, l, l_{0 \rightarrow N-2})\}$. These models can be plugged into Algorithm~\ref{algo1}. In the $K$th iteration of loop at line 3, we can employ $f_{K, \theta_{K}}(q, u, l, l_{0 \rightarrow K-1})$ to evaluate the logits, using listings already placed at $0$ through $K-1$ as antecedent listings. The logits are computed afresh in each iteration of the loop, now incorporating diversity. 

Though the theory presented in this section is simple, it is not a very practical one. For ranking $N$ positions, the number of models needed to be trained is $O(N)$. On top of it, the computational complexity of Algorithm~\ref{algo1} is $O(N^3)$, since the loop at line 2 is $O(N)$, the iteration over each listing at line 3 is $O(N)$, and the complexity of evaluating $f_{K, \theta_{K}}(q, u, l, l_{0 \rightarrow K-1})$ is $O(N)$. In the next section we discuss how to make the framework more practical.

\section{How to efficiently implement diversity in ranking?}\label{crt2}
We start with the $N$ distinct models constructed for each of the $N$ positions, and simplify them one by one.

The model for position $0$, $f_{0, \theta_0}(q, u, l)$, is our regular pairwise booking probability model $f_{\theta}(q, u, l)$ from Section~\ref{conehead}. We treat this as our core model and write the base case as:
\begin{equation}
 f_{0, \theta_0}(q, u, l) = f_{\theta}(q, u, l)
\end{equation}

To simplify $f_{1, \theta_1}(q, u, l, l_0)$, the model for position $1$, we compare it with $f_{\theta}(q, u, l)$. Both models are obtained by minimizing the cross-entropy loss over pairwise training examples consisting of booked and not booked listing pairs. The differences between them are:
\begin{compactitem}
\item $f_{\theta}(q, u, l)$ is trained over pairs constructed from all searches. $f_{1, \theta_1}(q, u, l, l_0)$ is trained on the subset of searches where the booked listing appears below position $0$.
\item $f_{\theta}(q, u, l)$ has the listing being ranked $l$ as the input, whereas $f_{1, \theta_1}(q, u, l, l_0)$ has the listing being ranked $l$, as well as the antecedent listing $l_0$, as inputs.
\end{compactitem}
We expect $f_{\theta}(q, u, l)$ and $f_{1, \theta_1}(q, u, l, l_0)$ to be fairly close since a large part of their training examples are shared. But we expect $f_{1, \theta_1}(q, u, l, l_0)$ to outperform $f_{\theta}(q, u, l)$ for position $1$ since it can downrank listings that are too similar to the antecedent $l_0$. We use this insight to simplify $f_{1, \theta_1}(q, u, l, l_0)$, refactoring it into two models as:
\begin{equation}\label{eq:sim}
f_{1, \theta_1}(q, u, l, l_0) = f_{\theta}(q, u, l) - s_{\phi}(q, u, l, l_0)
\end{equation}

First part of the refactor is $f_{\theta}(q, u, l)$, the regular pairwise booking probability model. The second part is a new model $s_{\phi}(q, u, l, l_0)$ parameterized by $\phi$. It adds a negative term based on the similarity between $l$ and $l_0$. This similarity is not {\it defined} by us, instead it is {\it learnt} from the training data that we built for $f_{1, \theta_1}(q, u, l, l_0)$. We train $f_{\theta}(q, u, l)$ prior to training $f_{1, \theta_1}(q, u, l, l_0)$. When training $f_{1, \theta_1}(q, u, l, l_0)$, we don’t need to train the $\theta$  parameters again. We can simply substitute $f_{\theta}(q, u, l)$ by the unconditional booking logit $ubl_l$, where $ubl_l=f_{\theta}(q, u, l)$ is obtained by evaluating the model for $l$. Thus,
\begin{equation*}
f_{1, \theta_1}(q, u, l, l_0) = ubl_l - s_{\phi}(q, u, l, l_0)
\end{equation*}

Given a training example for $f_{1, \theta_1}(q, u, l, l_0)$, with $l_x$ as booked, $l_y$ as not booked, and $l_0$ as antecedent, we have
\begin{equation*}
 \text{logit}_x=ubl_x - s_{\phi}(q, u, l_x, l_0) ; \text{logit}_y=ubl_y-s_{\phi}(q, u, l_y, l_0)
 \end{equation*}
 and cross-entropy loss defined by Equation~\ref{eq:crossentropy}. The parameters $\phi$ obtained by minimizing the cross-entropy loss over all the training examples gives us a simplified construction of $f_{1, \theta_1}(q, u, l, l_0)$ according to Equation~\ref{eq:sim}.

\begin{figure}
\includegraphics[height=1.45in, width=3.1in]{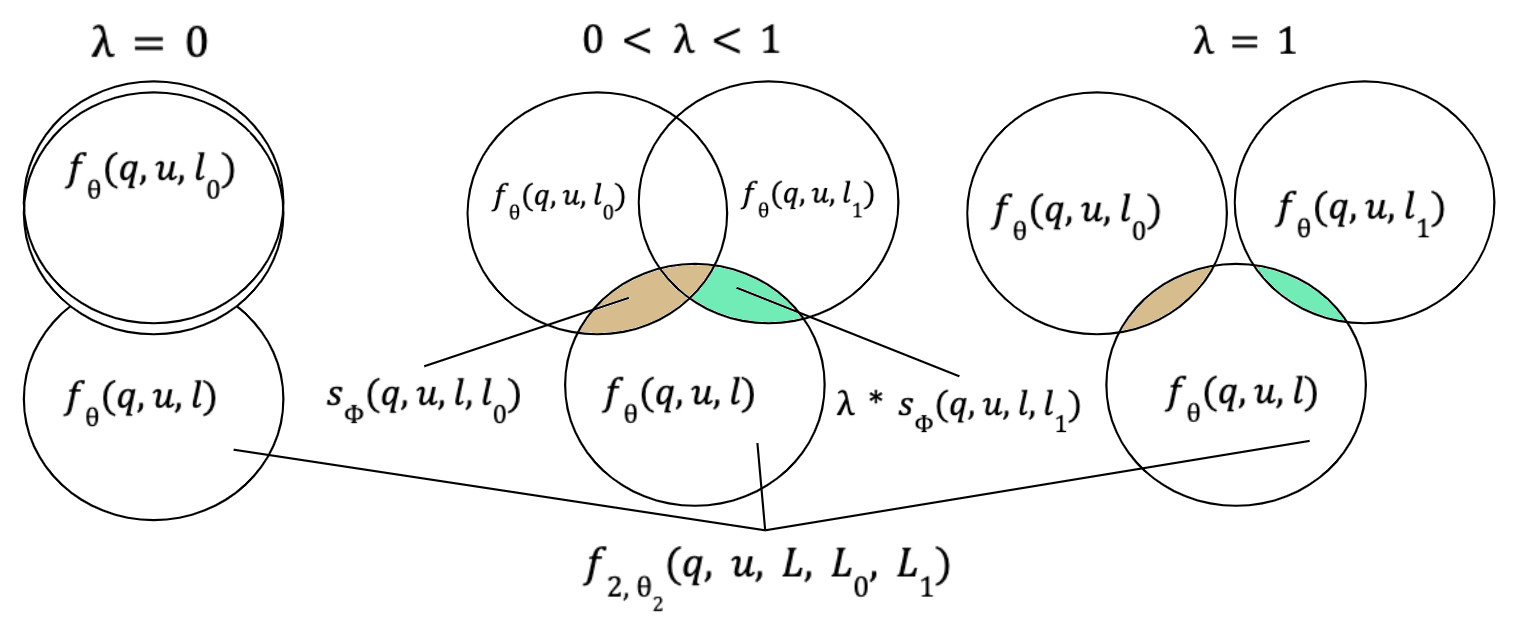}
\caption{\textmd{Illustration for Equation~\ref{eq:f23}}}
\label{fig:venndia}
\end{figure}

For position $2$, we have two antecedent listings, one at position $0$ and the other one at position $1$. We need to account for the similarity to both these antecedent listings. Though this time around, we do not need to learn the similarity model all over again. We can reuse the similarity model we learnt as part of $f_{1, \theta_1}(q, u, l, l_0)$. The refactored model for position $2$ can therefore be written as:
\begin{equation}\label{eq:f2}
 f_{2, \theta_2}(q, u, l, l_0, l_1) = ubl_l - s_{\phi}(q, u, l, l_0) - s_{\phi}(q, u, l, l_1)
 \end{equation}
But Equation~\ref{eq:f2} is valid only if the effect of $l_0$ and $l_1$ are completely independent of each other. On the other hand, if $l_0$ was an exact duplicate of $l_1$, then $l_1$ would not have any incremental effect and we could completely ignore it to rewrite Equation~\ref{eq:f2} as
\begin{equation}\label{eq:f22} 
f_{2, \theta_2}(q, u, l, l_0, l_1) = ubl_l - s_{\phi}(q, u, l, l_0)
\end{equation}
In reality, we expect the true effect to be somewhere in between Equation~\ref{eq:f2} and ~\ref{eq:f22}, which we write as
\begin{alignat}{2}\label{eq:f23}
\begin{split}
 f_{2, \theta_2}(q, u, l, l_0, l_1) &= ubl_l - s_{\phi}(q, u, l, l_0) - \lambda*s_{\phi}(q, u, l, l_1) \\
f_{2, \theta_2}(q, u, l, l_0, l_1) &= f_{1, \theta_1}(q, u, l, l_0) - \lambda*s_{\phi}(q, u, l, l_1)
 \end{split}
 \end{alignat}
where $0 \le \lambda \le 1$.
Figure~\ref{fig:venndia} depicts Equation~\ref{eq:f23} in terms of a Venn diagram.

For position $3$, we have to factor in the effect of similarity to the additional antecedent listing $l_2$ given by $s_{\phi}(q, u, l, l_2)$. But we expect the effect to be reduced under the assumption that $l_2$ isn’t completely independent of $l_0$ and $l_1$. To denote the incremental impact of $l_2$, we scale the contribution from $l_2$ by $\lambda ^ 2$, once for overlap with $l_0$, the second time for overlap with $l_1$. This gets us the refactored equation:
\begin{alignat*}{2}
f_{3, \theta_3}(q, u, l, l_0, l_1, l_2) =  f_{2, \theta_2}(q, u, l, l_0, l_1) - \lambda^2*s_{\phi}(q, u, l, l_2)
 \end{alignat*}
Generalizing for position $(k+1)$, we can write
\begin{alignat*}{2}
f_{k+1, \theta_{k+1}}(q, u, l, l_{0 \rightarrow k}) &= f_{k, \theta_{k}}(q, u, l, l_{0 \rightarrow k-1}) - \lambda^{k}*s_{\phi}(q, u, l, l_{k}) \\
   & = f_{\theta}(q, u, l) - \sum\limits_{i=0}^{k} \lambda^{i}*s_{\phi}(q, u, l, l_{i})
\end{alignat*}
We now need to build only two models, $f_{\theta}(q, u, l)$ and $s_{\phi}(q, u, l, l_0)$. The models for each of the positions can be expressed in terms of those two.

Using $f_{\theta}(q, u, l)$ and $s_{\phi}(q, u, l, l_0)$, we can construct Algorithm~\ref{algo2} which provides an iterative way to construct the search results, while taking diversity into account.

\begin{algorithm}
\caption{Ranking diversely} \label{algo2}
\begin{algorithmic}[1]
\Require{A set of $N$ listings $\{l_0, l_1, \dots l_{N-1}\}$}
\Ensure{Listing positions $\{\text{pos}(l_0), \text{pos}(l_1), \dots \text{pos}(l_{N-1})\}$}
\Statex
\State $\mathcal{L}  \gets \{l_0, l_1,..., l_{N-1}\}$
\State Compute $\text{logit}(l_i) \gets f_{\theta}(q, u, l_i)$ foreach $l_i \in \mathcal{L}$
\State $l_{max} \gets \text{argmax}(\text{logit}(l_i), l_i \in \mathcal(L))$
\State $\text{pos}(l_{max}) \gets 0$
\State $l_{\text{atcdnt}} \gets l_{max}$
\State $\mathcal{L} \gets \mathcal{L} \setminus l_{max}$
\For {$k \gets 1$ until $N$}
    \State $\text{logit}(l_i) \gets \text{logit}(l_i) - \lambda^{k}*s_{\phi}(q, u, l, l_{\text{atcdnt}})$ foreach $l_i \in \mathcal{L}$
    \State $l_{max} \gets \text{argmax}(\text{logit}(l_i), l_i \in \mathcal(L))$
    \State $\text{pos}(l_{max}) \gets k$
    \State $l_{\text{atcdnt}} \gets l_{max}$
    \State $\mathcal{L} \gets \mathcal{L} \setminus l_{max}$
\EndFor
\end{algorithmic}
\end{algorithm}

We can treat $\lambda$ as a hyperparameter here and sweep through different values to find the one that maximizes $NDCG$. In our case, we settled at $\lambda = \frac{1}{3}$.

The simplification in this section reduced the number of models from $O(N)$ to $2$, and the computational complexity from $O(N^3)$ to $O(N^2)$. Line 8 is now the bottleneck in Algorithm~\ref{algo2}. The number of iterations of the loop is $O(N)$, and iterating over each listing in line 8  is $O(N)$. The main computation in this inner loop is evaluation of $s_{\phi}(q, u, l, l_{\text{atcdnt}})$. We can further reduce the complexity by optimizing the evaluation of $s_{\phi}(q, u, l, l_{\text{atcdnt}})$.

\begin{figure}
\includegraphics[height=2.2in, width=3in]{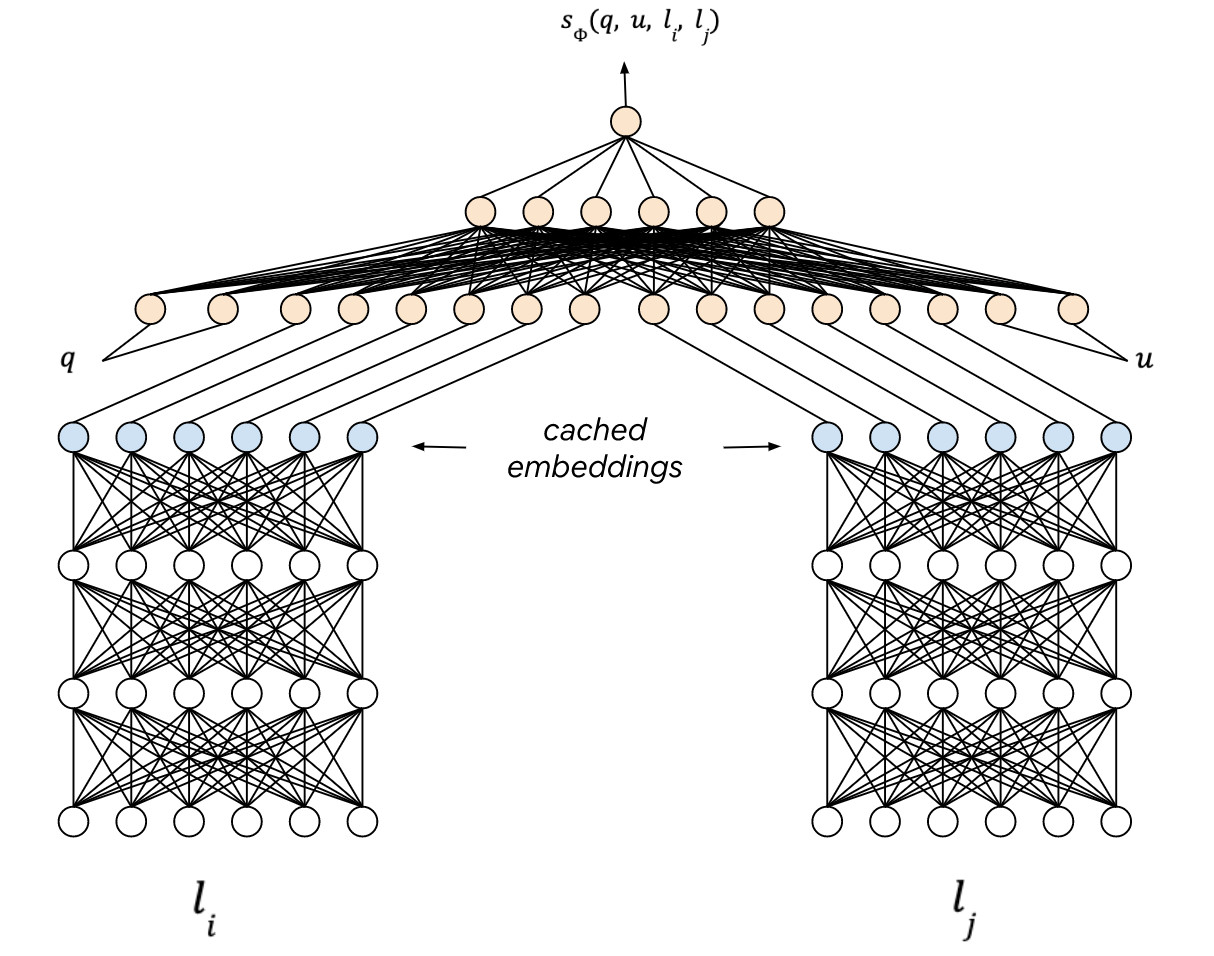}
\caption{\textmd{Neural network for $s_{\phi}(q, u, l_i, l_j)$. Weights are shared for the listing towers and their output is cached.}}
\label{fig:simtower}
\end{figure}

Our discussion until this point did not assume any model architecture. But for optimizing the evaluation of $s_{\phi}(q, u, l, l_{\text{atcdnt}})$, we limit the discussion to neural networks, which is the model we implemented. The bulk of the model complexity comes from processing of the listing features. We use a tower of fully-connected layers to map the listing features into an embedding. A shallow layer at the top then combines the embeddings of the two input listings to finally output the similarity logit. See Figure~\ref{fig:simtower}.

The advantage of this architecture is that the output of the listing towers can be cached. We need to process each listing only once to map them to their corresponding embeddings. In line 8 of Algorithm~\ref{algo2}, we then reuse the cached embeddings and only evaluate the top part of the network. With these optimizations we were able to keep the latency impact to a minimum.

\section{How does this compare to other existing approaches?}
\subsection*{Enforcing Fairness in Ranking}
Imposing fairness constraints has been a popular method for diversifying search results. A ranking model to optimize a utility such as $NDCG$, within a specified upper bound of unfairness is discussed in ~\cite{policyfairness}. An alternate approach to provide different groups equal impressions or proportional impressions in web search results is presented in ~\cite{fairerranking}. Search results are built iteratively, where each step of the algorithm either chooses to optimize fairness with probability $\varepsilon$, or relevance with probability $1-\varepsilon$. The method in ~\cite{fairerranking} is attractive because of its simplicity, as it does not require building any new models.

The limitation of these approaches is that they inherently pose the problem as a tradeoff between utility and fairness. Utility is the quantifiable benefit in the short term, bookings in our case, which needs to be sacrificed to accommodate fairness, with assumed benefits in the long term. These benefits of fairness are {\it assumed} because they are hard to quantify given the long time horizon needed for tests. As a result, the knob between utility and fairness has to be controlled by faith. The strength of approaches such as ~\cite{policyfairness} and ~\cite{fairerranking} is that if one has a fairness goal in mind for which one is willing to sacrifice utility, such as gender equality between job applicants, or equality between opposing views on a topic, then these methods can achieve fairness while minimizing the utility loss. This is something our framework cannot handle, which can diversify only in the direction that aligns with optimizing utility. This exposes a fundamental difference between two goals: 1) imposing some fairness constraints on search results, and 2) providing users with the optimal level of choice in search results. Both goals on the surface are related to diversification, but in the final analysis, they lead to different paths.

\subsection*{Providing Context Features}
Another popular method to diversify ranking is to provide features of surrounding items in search results. These approaches are closer to our framework as they aim to improve utility through diversification, and diversity is not a goal in itself. An example is the ranking of widgets on Amazon video homepage, discussed in ~\cite{amzndiversity}. The method relies on a predefined categorization of widgets based on a combination of content type and purchasing option. Features are then constructed to summarize the categories of the widgets placed in positions $0$ through $k-1$. These features capture the incremental gain in category diversity while scoring widgets for the $k$th position. The limitation of this approach is that it diversifies only along the predefined categories. In the case of Airbnb, we found such categorization of listings quite challenging. Attempts to diversify listings based on location, price, amenities, etc. mostly lead to negative results. When categorizing based on any particular criteria, for instance percentile price buckets, one ends up ignoring the rest of the dimensions such as location, quality, amenities, and aesthetics. Listings may bubble up in ranking simply because they belong to a certain price percentile bucket, compromising on one or more of the other dimensions that users care about. To prevent that, diversity needs to account for the entire context of the user, query, and listings -- and the interactions between them. Our method aims for this generalized diversification. The strength of ~\cite{amzndiversity} is its simplicity. If there is a categorization available at hand, and one cares about diversity only along those categories, then ~\cite{amzndiversity} reduces the problem to feature engineering and does not require any new model to be learnt.

Our previous attempt at diversity described in ~\cite{kdddiversity} provides another method for supplying context information to the model. A recurrent neural network is used to map the features of all the listings in the search result to an embedding. This embedding can then be used as a feature by the ranking model to make more context aware decisions. The challenge with this method is loss of information. While ranking for the $k$th position, the embeddings don't provide information about the listings placed at $0$ through $k-1$. Instead, the embeddings represent an aggregated summary of all the listings mushed together, and doesn’t allow the kind of pairwise comparisons a user would perform. The strength of this technique is its simplicity when scoring. Once the embedding is evaluated, plugging it in as a feature is straightforward, and requires no additional computation. 

\subsection*{Sequential and Setwise Ranking}
Abandoning the assumption that utility of individual items in search results are independent of each other, and making ranking aware of their interactions has been gaining momentum in recent years. A reinforcement learning algorithm is used in ~\cite{mdpdiversity}, where ranking for each position is considered one time step of the sequential Markov decision process. While ranking for the $k$th position, the state of the RL algorithm encodes the items placed at positions $0$ through $k-1$. At each step, a Monte Carlo tree search is used to explore the updates to policy. The challenge with this approach is its complexity, both for training the RL algorithm and for serving it online. Our approach reuses the pairwise booking probability model for the most part, combining it with a similarity model that is further optimized to reduce the computational complexity. Even then, we found a significant impact on online latency as we discuss under experimental results. The strength of ~\cite{mdpdiversity} is that it explores the space of possible rankings more thoroughly, and could potentially find larger $NDCG$ gains.

Taking all the items to be ranked as inputs and producing position assignments simultaneously is discussed in ~\cite{Pang2020SetRankLA}. The challenge with this approach, once again, is its complexity. In our setting each listing has \textasciitilde$700$ features, along with \textasciitilde$200$ features for the query and the user. For diversification $O(100)$ listings are evaluated. At this scale, the proposed setwise architecture is impractical given the system constraints for training and serving. The strength of the setwise approach is that it makes the least assumptions, and can be used as a tool to assess the upper bounds of the gains possible through diversification. A method to tackle the complexity of setwise ranking is discussed in ~\cite{youtube}, which uses determinantal point processes. It is an alternative to our proposal that also relies on the concept of similarity.

Direct optimization of the diversity aware metric $\alpha\text{-}NDCG$ through a differentiable diversification-aware loss is discussed in ~\cite{alpha_ndcg}. In comparison, our approach shows how diversity can be achieved through optimization of regular $NDCG$ itself.

\subsection*{Listwise Loss}
We previously discussed how typical pairwise learning to rank frameworks consider only the attributes of the listing being ranked. Listwise learning to rank considers the entire list, so presumably they can overcome that limitation. Even then, in common listwise learning to rank algorithms, such as ListNet ~\cite{listloss}, it is assumed that the booking probability of individual listings are independent of each other. This assumption is necessary to keep the computational complexity tractable. For handling diversity, it is not enough to have access to all the other listings in the search result, one has to explicitly take into consideration the interaction between the listings. Accounting for the interactions leads to the kind of problem formulation described in Section~\ref{crt}.

The assumption regarding the independence of items ranked is relaxed in ~\cite{listcontext}. It encodes the features of the top results into an embedding using a recurrent neural network, which are then used as supplemental features for scoring each item. The comparison of ~\cite{listcontext} to our work is similar to what we discussed for ~\cite{kdddiversity}.

\section{How does the theory work in practice?}
We developed the theory in Sections~\ref{crt} and ~\ref{crt2} before embarking on actual implementation and experimentation. This allowed us to make certain predictions about the experimental results. In a flow of events reminiscent of developments in physics, the predicted experimental results were subsequently matched by tests offline and online. 

\subsubsection*{NDCG}
The first prediction to come out of the theory was a simple one: that $NDCG$ should improve. What made the prediction interesting was the expectation of an $NDCG$ gain, with no additional information added to the training data. In contrast, most $NDCG$ gains over the past years required incrementally new information in the form of features, labels, or bug fixes. On the test set for $f_{1, \theta_1}(q, u, l, l_0)$, we observed an $NDCG$ gain of $0.45\%$ compared to the baseline Algorithm~\ref{algo1}. When measured on the test set of $f_{\theta}(q, u, l)$, this gain translated to $0.2\%$. This gain is smaller than $0.45\%$ because the listing at the first position remains invariant between Algorithm ~\ref{algo1} and ~\ref{algo2}, diluting the effect. The impact on $NDCG$ measured in the online A/B test was much stronger, where we observed an increase of $1.5\%$.
\subsubsection*{Bookings \& Booking Value}
From the $0.2\%$ $NDCG$ gain, we expected a similar gain in bookings online. Further, we expected these bookings to come from preference groups which were a minority in the training data. The most prominent minority preference being quality-leaning searchers, as discussed in Section~\ref{conehead}.
This allowed us to make a further prediction: that there would be gains in gross booking value, or the sum of the prices of the booked trips, which would be multiple times over the bookings gain. In the online A/B test, we observed a bookings gain of $0.29\%$. Along with it, we saw a $0.8\%$ gain in gross booking value. Segmenting the bookings gain by user groups, we found almost the entire gain came from users who were booking a listing on Airbnb for the first time.
\subsubsection*{Engagement}
Other observations from the online A/B test include a $0.46\%$ increase in listings viewed, and a $1.1\%$ increase in listings saved. This increased engagement with search results can be attributed to the increased choice.
\subsubsection*{Price \& Location}
To directly measure diversification, we compared metrics along some key dimensions. The first measure compared the variance in price among the top $8$ results. We observed an increase of $3.4\%$ in treatment, which captures the increased diversity in price, and hence quality by proxy. The second metric compared the number of listings in the top $8$ results that were within $0.5$ km of each other. We noted a decrease of $0.62\%$, which shows reduced redundancy in location.
\subsubsection*{Trip Quality}
To measure the impact of diversity on the entire user experience, we waited for $90$ days after the end of the A/B test. This allowed the majority of the trips booked during the experiment to be realized. Comparing the ratings from guests checking out of their stays, we noted an increase of $0.4\%$ in 5-star ratings. Diversifying search results shifts the balance away from the majority preference of affordability towards the minority preference of quality. This shift towards quality ultimately surfaces in improved trip ratings.
\subsubsection*{Latency}
The gains came at a latency cost, where we saw an increase of $8.4\%$ in P95 latency and a $5.3\%$ increase in median latency.
 
\section{Conclusion}
We started this paper discussing how ranking evolved at Airbnb. We conclude with a summary of our efforts to diversify ranking. Early attempts started in 2017 with category-based diversification. Various categories were tried, based on price, location, and amenities. All these efforts resulted in disappointment. This led to the conclusion that diversifying along a particular dimension mostly degraded the quality of results. Focus then moved to diversification along multiple dimensions, in particular combining price and location. These attempts resulted in failure as well. The breakthrough came in 2019 with ~\cite{kdddiversity}, where instead of forcing ranking to adhere to some predefined notion of diversity, we supplied the ranking model with more information, giving it the freedom to diversify. The current work continues with that philosophy. We revisited the problem in 2022 with a theory-first approach, letting the model learn the notion of diversity from the training data. This lead to one of the most impactful ranking changes of the year. But as discussed in Section~\ref{conehead}, this training data itself is biased against diversity. After the launch of the diversity ranker, we expect future training data to have richer examples to learn from, enabling a virtuous cycle of diversification.

%\end{document}  % This is where a 'short' article might terminate

%\bibliographystyle{ACM-Reference-Format}
\bibliographystyle{ACM-Reference-Format}

\balance 
\bibliography{diversity-bibliography}

\end{document}